\documentclass{emulateapj}
\begin{document}
\newcommand{\etal}{\mbox{et al.}}
\newcommand{\acm}{ cm$^{-2}$ }
\newcommand{\pp}{\phn}
\newcommand{\ppp}{\phn\phn}
\newcommand{\pppp}{\phn\phn\phn}
\newcommand{\pq}{\,$\pm$\,}
\newcommand{\pd}{\phn\phn\phn\,---}
\newcommand{\plong}{\hspace{10pt}}
\newcommand{\Msun}{M$_{\odot}$}
\newcommand{\ayr}{y$^{-1}$}
\newcommand{\LCDM}{$\Lambda$CDM}
\newcommand{\ARAA}[2]{ARA\&A, #1, #2}
\newcommand{\ApJ}[2]{ApJ, #1, #2}
\newcommand{\ApJL}[2]{ApJL, #1, #2}
\newcommand{\ApJSS}[2]{ApJS, #1, #2}
\newcommand{\ApJinp}{ApJ, in press}
\newcommand{\ApJsub}{ApJ, submitted}
\newcommand{\AandA}[2]{A\&A, #1, #2}
\newcommand{\AandASS}[2]{A\&AS, #1, #2}
\newcommand{\AJ}[2]{AJ, #1, #2}
\newcommand{\BAAS}[2]{BAAS, #1, #2}
\newcommand{\ASP}[2]{ASP Conf.\ Ser., #1, #2}
\newcommand{\JCP}[2]{J.\ Comp.\ Phys., #1, #2}
\newcommand{\MNRAS}[2]{MNRAS, #1, #2}
\newcommand{\MNRASinp}{MNRAS, in press}
\newcommand{\MNRASsub}{MNRAS, submitted}
\newcommand{\N}[2]{Nature, #1, #2}
\newcommand{\PASJ}[2]{PASJ, #1, #2}
\newcommand{\PASP}[2]{PASP, #1, #2}
\newcommand{\PRD}[2]{Phys.\ Rev.\ D, #1, #2}
\newcommand{\PRL}[2]{Phys.\ Rev.\ Lett., #1, #2}
\newcommand{\RPP}[2]{Rep.\ Prog.\ Phys., #1, #2}
\newcommand{\ZA}[2]{Z.\ Astrophs., #1, #2}
\newcommand{\tenup}[1]{\times 10^{#1}}
\newcommand{\pz}[0]{\phantom{0}}

\title{Mass Profiles of Galaxy Cluster Cores:\\
Implications for Structure Formation and\\
Self-Interacting Dark Matter}
\author{J.S.\ Arabadjis\altaffilmark{1} and M.W.\ Bautz\altaffilmark{1}}
\altaffiltext{1}{Center for Space Research, Massachusetts Institute of
Technology, Cambridge, MA 02139; {\tt jsa@space.mit.edu},
{\tt mwb@space.mit.edu}}

\begin{abstract} 

We present a spectroscopic deprojection analysis of a sample of ten relaxed
galaxy clusters.  We use an empirical F-test derived from a set of Markov Chain
Monte Carlo simulations to determine if the core plasma in each cluster could
contain multiple phases.  We derive non-parametric baryon density and
temperature profiles, and use these to construct total gravitating mass
profiles.  We compare these profiles with the standard halo parameterizations.
We find central density slopes roughly consistent with the predictions of
\LCDM: \mbox{$-1 \lesssim d\log(\rho)/d\log(r) \lesssim -2$}.  We constrain the
core size of each cluster and, using the results of cosmological simulations as
a calibrator, place an upper limit of
$\sim 0.1$ cm$^2$ g$^{-1}$ = 0.2 b(GeV/c$^2$)$^{-1}$ (99\% confidence) on the
dark matter particle self-interaction cross section.

\end{abstract}


\section{Introduction} 

The \LCDM\ theory of cosmological structure formation -- a cold dark matter
(CDM) universe evolving under gravitation with a cosmological constant
($\Lambda$) -- has been very successful in characterizing large scale structure
on scales greater than $\sim 0.1$ Mpc \citep{mohr,lahav,sperg_etal,perc_etal}.
However, \LCDM\ makes three predictions which appear to conflict with
observations.  The first of these discrepancies is the ``satellite problem'' --
the Milky Way contains at least an order of magnitude fewer satellites than
simulations predict for a dark matter halo of its size
\citep{kauffman,moore_a,klypin}.  The second is the ``disk problem,'' -- disk
galaxies produced in simulations contain too little mass and angular momentum
\citep{navarro}.  The third is the ``core problem'' -- galaxy-sized dark matter
halos produced in simulations tend to have significantly steeper logarithmic
density gradients in their cores than do observed structures.

The density profile of a dark matter halo formed through the hierarchical
assembly of smaller structures is usually parameterized as a pair of power
laws with a transition scale radius $r_s$ (e.g.\ \citet{jing}),
\begin{equation}
\frac{\rho(r)}{\rho_0} = \frac{1}{(r/r_s)^\alpha \, (1+r/r_s)^{\gamma-\alpha}}
\, ,
\label{eq01}
\end{equation}

\noindent with $\alpha = 1$ (\citet{nfw_six,nfw_seven}; hereafter the ``NFW''
profile) or $\alpha = 1.5$ (\citet{moore_b}; hereafter the ``Moore'' profile).
While there is general agreement that $\gamma = 3$, rotation curves of low
surface brightness \citep{burkert,mcgaugh} and spiral galaxies
\citep{salucci,gentile} and the fundamental plane of elliptical galaxies
\citep{borriello} suggest that core density profiles are significantly flatter
than \LCDM, with $0 \geq \alpha \geq 1$.

Reports of a core problem on galaxy cluster scales have been somewhat more
controversial.  Some lensing studies have found flat core profiles
\citep{tyson,smail} while others, making use of similar data sets, have not
\citep{broadhurst,shapiro}.  Most recently, \citet{sand} and \citet{sand_new}
have found shallow ($\rho \sim r^{-0.5}$) cores in several clusters using a
combination of radial and tangential strong lensing arcs in tandem with stellar
velocity dispersion measurements.

The severity and cause of these discrepancies have been a subject of intense
debate for nearly a decade.  While observational limitations may play a role
\citep{swaters_newest}, they are unlikely to be the sole cause in the majority
of cases \citep{dalcanton,swaters,smvdbb,bolatto}.  Various explanations have
been offered which can be ascribed to astrophysics or to particle physics.
Most \LCDM\ simulations evolve collisionless dark matter particles and baryons
under the influence of gravitation only; final baryon distributions are usually
determined according to prescription afterward.  Thus, the disagreement between
simulation and observation is either caused by neglected baryonic processes
(astrophysics) which could act to flatten the baryon-dominated cores, or could
be the result of neglected processes involving dark matter (particle physics).

It is certainly plausible that many of the problems on scales of $\sim10$ kpc
are the result of neglected baryon physics, if only because this is the scale
at which star formation and the associated processes dominate the energy
budget in galaxies.  It seems less likely that discepancies over the range
100-800 kpc, scales probed in the \citet{sand_new} study, could have the same
provenance.  Thus if these measurements are correct they may require a
significant revision of our understanding of structure formation and the
behavior of dark matter.  The literature already provides a plethora of
alternative dark matter theories:  self-interacting dark matter
\citep{sperg_stein}, warm dark matter \citep{hogan}, annihilating dark matter
\citep{kaplinghat}, scalar field dark matter \citep{goodman,hu}, decaying dark
matter \citep{cen}, and mirror matter \citep{mohapatra,foot}, to name several.
Of these, self-interacting dark matter (SIDM) \cite{sperg_stein} has probably
garnered the most attention, since with only one or two scattering events per
particle over the age of the universe the cores of galaxies could be
sufficiently softened to match observed galaxy profiles \cite{dssw} (hereafter
DSSW).  SIDM is also theoretically attractive in that it would produce softened
cores on galaxy cluster scales \cite{yswt}, potentially providing an
explanation for the \citet{sand_new} results.

The $10^8$ K X-ray emitting cluster plasma provides an important alternative
probe to gravitational lensing reconstructions of cluster mass profiles.
Lensing studies, which measure projected mass, may preferentially select
clusters with significant substructure along the line-of-sight \citep{cohn}.
In addition, most lensing-derived masses are model-dependent reconstructions
(e.g.\ \citet{sand_new}).  On the other hand, while X-ray mass determinations
can utilize non-parametric spectroscopic deprojection, they are valid only if
the hot plasma is in hydrostatic equilibrium.  Those clusters for which this is
most likely true usually display ``cooling flow'' emission which can alter the
derived profile depending on how it is modelled (\citet{allen,abg,sksmec,aba},
hereafter ABA).  Since these methods employ completely independent assumptions
and potentially suffer from different systematic uncertainties, they constitute
a powerful two-pronged attack on the core problem.

It has been argued \citep{markevitch_he} that the complex cluster surface
brightness distributions revealed by Chandra throw the hydrostatic hypothesis
into doubt.  We take a more optimistic attitude, noting that we select clusters
which are largely devoid of significant substructure.  In addition, our method
includes a diagnostic for clusters whose cores are not in hydrostatic
equilibrium (see \S\ref{technique}).  We note that \citet{kay_etal} find that
neglecting a kinematic pressure term from a hydrostatic analysis leads to
errors in the determination of cluster masses of only $\sim 10$\% for those
whose cores contain cool plasma.

In this paper, we use X-ray imaging spectroscopy from the Chandra X-ray
Observatory to constrain cluster mass profiles under the assumption of
hydrostatic equilibrium.  We briefly outline our mass profile extraction
technique in \S\ref{technique}.  We then examine the core mass profiles of a
sample of relaxed clusters and compare with the predictions of \LCDM\ in
\S\ref{LCDM}.  Finally we use these profiles to constrain the self-interaction
cross section of dark matter particles in \S\ref{SIDM}.

\section{Extraction of mass profiles} \label{technique} 

The spectroscopic deprojection and mass reconstruction technique we employ has
been fully described elsewhere \citep{abg,aba} so we only briefly summarize
them here.  Starting with a Chandra ACIS-S or ACIS-I observation of a cluster,
we process the data in the usual
way\footnote{See http://asc/harvard.edu/ciao/threads/all.html}, removing point
sources and periods of high background.  We divide the events into concentric
annuli centered on the emission peak such that there are at least 1000-1500
source photons per annulus.  We construct a model consisting of a set of
concentric spherical shells whose inner and outer radii $r_i$ and $r_o$
correspond to the angular radii of the annuli in the data.  Each shell is
modelled as a thermal plasma at redshift $z$ using the MEKAL model
\citep{meweg,mewel,kaastra,liedahl} within XSPEC \citep{arnaud}.  Each plasma
component is characterized by a temperature $T$ and normalization $K$.  (The
metal abundances are set to 0.3 solar.)  The normalization is related to the
plasma density $\rho = \mu \, m_p \, n_H$ through the shell geometry,
\begin{equation}
K^{-1} = 10^{14} \, [4\pi \, D_A (1+z)]^2 \, \int_{r_i}^{r_o} \, n_e \, n_H \,
r^2 \, dr
\label{eq02}
\end{equation}

\noindent where $D_A$ is the angular diameter distance of the cluster.

We test each cluster for the presence of a second (cooler) cospatial emission
component in the core using an empirical F-test applied to a suite of 1000
Markov Chain Monte Carlo (MCMC) simulations per cluster.  We use the
MCMC significance $S$ -- defined as one minus the fraction of simulations which
show an improvement in the model fitting as large as that shown by the data
simply by chance -- to decide whether a second core emission component is 
required, adopting a threshold of 0.99 for its inclusion (ABA).

We select clusters from the Chandra archive which display (i) a single peak in
their X-ray emission and circular isophotes, and (ii) no obvious signs of
merging activity or other asymmetry in their X-ray surface brightness
distribution.  Since these criteria are somewhat subjective, we further screen
clusters in the sample according to the shape of their core mass profiles.
Specifically, when deriving ensemble averages, we use only those clusters for
which a power law is an acceptable fit ($\chi^2_r<1.3$) to the core profile
(the innermost 5 points in the profile, usually $r \lesssim 100$ kpc).  We
shall henceforth refer to these as the ``relaxed core cluster'' (RCC) subset.
This prevents contamination of our sample from clusters which appear relaxed
but whose irregular core mass profiles may belie the hydrostatic hypothesis.
Seven of the 10 clusters in our original sample (ABA) are RCCs, though we
include the remaining three objects for comparison in parts of this analysis.
Table~\ref{t01} lists the 10 clusters in the sample, their redshift, the
treatment adopted for their core emission, the MCMC significance of the second
component, and whether or not the cluster core is relaxed.

The model parameters, $T_i$, $K$ ($i=1,N$), and a Galactic absorption column
$N_{\rm HI}$ are fit for simultaneously, and 2D confidence limits for each
$T_i,K_i$ (alternatively $T_i,\rho_i$) pair are calculated.  The runs of
temperature and density for each cluster are then used to calculate the total
enclosed gravitating mass of the cluster $M_r$ using the hydrostatic equation
\citep{sarazin}:
\begin{equation}
M_r = 3.535 \times 10^{10} \, {\rm M}_{\odot} \,
\left(\frac{T}{{\rm 1 \, keV}}\right) \,
\left(\frac{r}{{\rm 1 \, kpc}}\right) \,
\left( -\frac{d\log{T}}{d\log{r}} -\frac{d\log{\rho}}{d\log{r}} \right)
\label{eq03}
\end{equation}

\noindent Statistical fluctuations in the temperature measurements can result
in an unphysical mass determination, i.e.\  $M_r<0$ or $dM/dr<0$, over a
limited range in $r$.  To remedy this situation we {\it impose} the constraints
$M_r>0$ and $dM/dr>0$ on the mass profile and allow the temperature and density
of each shell to vary slightly in order to satisfy them.  The degree to which
these constraints are enforced is a free parameter (i.e.\ a weighting factor in
a penalty function); if the variation in $T$, or less likely $\rho$, required
to satisfy the constraints is excessive (e.g.\ $\Delta T/\sigma_T \gtrsim 1$)
we conclude that one of our input assumptions, and therefore the mass profile,
is suspect (ABA).

\section{Core profiles} \label{LCDM} 

In Figure~\ref{f01} we show mass profiles (top) and cumulative power law slopes
$\alpha$ (bottom) for all 10 sample clusters, with uniphase models at left
and two-phase models at right.  The exception to this is Hydra~A, where we
show only a uniphase model.  Our attempts at fitting a second core component
in this cluster resulted in its best-fit temperature matching that of the
original component, consistent with \citet{david}.  For each cluster, the model
with the higher MCMC F-test significance (ABA; see also Table~\ref{t01}) is
indicated by the `preferred' label.

The cumulative power law slope, plotted as filled squares, represents a power
law fit to the profile as a function of the outermost point used in the
fitting.  The innermost value of $\alpha$ is duplicated at the minimum $r$
value with an open square for clarity (except for Hydra A and A1795, whose core
profiles are unphysical even with a significant weighting of the mass
constraints term in the penalty function).  \LCDM\ predicts that $\alpha$
should asymptotically approach $-1$ (the NFW profile) or $-1.5$ (the Moore
profile) as $r$ decreases.

\subsection{Uniphase versus two-phase} \label{oneortwo} 

In general two-phase models have steeper central mass profiles than do uniphase
models.  This is so because the two-phase models allow the temperature of the
hot component to remain roughly constant all the way in to spatial resolution
limit.  This augments the mass at each radius (see Equation~\ref{eq03}), making
$M_r$ flatter in the center.  Since $M_r \sim r^{\alpha+3}$, this implies a 
steeper density profile.

\subsection{Comparison with previous studies} \label{comp} 

The profiles of power law slope show variation among the sample clusters, and
in some cases between different plasma emission models (Figure~\ref{f01}).
The core region of two of the clusters in our sample -- Hydra A
\citep{mcnamara,david}, A1795 \citep{markevitch_ab,ettori}, are known to be out
of hydrostatic equilibrium.  Figure~\ref{f01} shows that the inferred mass
profiles for these clusters are indeed irregular.  Three of the most relaxed
clusters in our sample -- A2029, A1689 and MS2137 -- have been modelled in
detail in other studies, allowing us to compare mass profiles.

\subsubsection{A2029 } \label{a2029} 

The inferred core profile of A2029 is sensitive to the assumed plasma model
(Figure~\ref{f01}c).  The uniphase model has a nearly flat core, whereas the
two-phase core, which is preferred $> 99$\% significance, is consistent with
the Moore profile.  Confidence contours in $\alpha$ and the normalization $M_0$
(where $M_0 = M_r(1{\rm Mpc})$) are shown in the upper left panels of
Figure~\ref{f02}.  The preferred two-phase model shows a core density that is
even steeper than the Moore profile.  These results differ from those of
\citet{lewis_b}, who find no evidence for multiphase core plasma, based on a
lack of significant reduction in the $\chi^2$ value of their spectral fits, and
\citet{lewis_a} (hereafter LBS), who find that NFW provides a good fit to their
mass profile.  These discrepancies may be due in part to the specifics of the
deprojection method used in each case.  Our technique requires more photons per
annulus than the ``onion peeling'' approach, essentially because it is
non-parametric and global, to adequately constrain the temperature at each
radius.  We must therefore use larger regions for the extraction of spectra.
The result is that we resolve temperature structure down to about 15 kpc,
whereas LBS resolve it to $\sim 3$ kpc.  In their model the plasma temperature
drops a factor of two over this range, suggesting that perhaps the strong
two-phase signature we see is due to a unresolved core temperature structure.
We find strong evidence, however, for a second plasma component out to 34 kpc,
well beyond our (and their) resolution limit, whereas LBS find none, so this
cannot be the sole cause of the disparity.  We note that \citet{clarke} also
find evidence for a second (cooler) emission component within $r<40$ kpc.

In addition to spatial resolution, a fundamental difference between these two
studies is the metric used to gauge the significance of the second component.
LBS conclude that there is no evidence for  multiphase plasma based on the lack
of significant reduction in the $\chi^2$ value of their fit \citep{lewis_b}.
Our method adopts the recommendation of \citet{protassov}, who have studied
statistical tests used to compare pairs of models when one lies on a
parameter-space boundary of the other (in our case, the normalization of the
second component set to zero).  The have found that the standard statistical
tests can underestimate the signal in noisy data.  In such cases one should use
a statistical test performed on an empirical distribution derived from a large
sample of data realizations which follow the complete parameter likelihood
function.  Following \citet{protassov}, our value for the significance of the
second component is derived from the F distribution of a large MCMC sample (see
\S\ref{technique}).  We believe therefore that our value for the the core slope
of A2029 is more reliable.

\subsubsection{A1689 } \label{a1689} 

\citet{andersson} used XMM-Newton data and the deprojection scheme of ABG
to derive a core slope of $-1.27$, which is consistent with our results.  It
should be noted that they find evidence for an asymmetry in the temperature
structure of the plasma, and suggest that this could be due to a merger event.
Our mass profile shows no evidence of a merger, although our model does
show a temperature jump at this distance (ABA, Figure~5).  It is not clear that
this affects the value obtained for the core slope, however.

\subsubsection{MS2137} \label{ms2137} 

Our profile for MS2137 is radically different from the strong gravitational
lensing results of \citet{sand}.  These authors use radial and tangential arcs
in conjunction with the stellar velocity dispersion profile of the central
galaxy to constrain the central density profile, finding $\alpha > -0.9$ at the
99\% confidence limit.  Our analysis yields $\alpha = -1.6\pm 0.2$ ($1\sigma$),
with $\alpha<-1.2$ at the 99\% CL.

This discrepancy {\it cannot} be due to the difference between deriving a dark
matter profile profile \citep{sand} and a total matter profile (this work).
Baryons dominate the mass profile only up to 10 kpc \citep{sand_new}; the
innermost of our mass measurements is at $r\sim 20$ kpc.  In addition, the
\citet{sand} analysis used a fixed NFW scale radius of 400 kpc; that is, 400
kpc marks the separation between the inner and outer power laws.  Since all of
our mass measurements are at radii smaller than this, our profile probes
precisely the range where \citet{sand} find a shallow inner profile.

We believe this is due to the nature of their lens model.  \citet{bartelmann}
have shown that flat cores are indeed required for spherical and axisymmetric
mass distributions, but if small deviations from axisymmetry are introduced
(such as a mild ellipticity in the lensing potential, or a small external
shear), these same systems are consistent with the cuspy profiles seen in
simulations.

\subsection{Central density slopes and \LCDM} \label{central} 

The confidence contours in Figure~\ref{f02}, shown for the RCC subset, show a
range of values for the central density slope.  This can be seen at the bottom
of Figure~\ref{f04}; we find that $-1 \le\alpha\le -2$. This distribution
appears somewhat steeper than CDM predictions (the gray band).  Using
symmetrized errors, we find a weighted mean of
$\langle a\rangle_{w} = -1.685 \pm 0.077$.  The fact that this is marginally
steeper than the CDM band is not surprising, since the contraction of cooling
baryons is expected to deepen the central cluster potential, and thus steepen
the central density profile (\citet{hennawi,gnedin}; but see also
\citet{loeb}).  None of the clusters in the sample is consistent with the flat
or nearly-flat cores found by \citet{sand_new}.  We note, however, that only
one cluster -- MS2137 -- appears in both samples.

One can interpret these results in at least two ways.  It is thought that
baryonic processes could act to steepen the mass profile as energy is removed
from the cluster core via relaxation and radiative cooling effects.  The
adiabatic contraction of the core baryons deepens the potential well, causing
the dark matter profile to steepen \citep{hennawi}.  Another possibility is
that merger tree hysteresis in the formation and evolution of the cluster,
distinct from baryonic physics, dictates the current value of slope of the core
dark matter profile \citep{ma}.  The question of which influence -- adiabatic
contraction or accretion hysteresis -- is dominant in setting the core profile,
and the degree to which heating from a central AGN may play a role
\citep{omma_a,omma_b}, may remain unsettled until numerical experiments are
able to self-consistently simulate the gas dynamics and cooling in conjunction
with the gravitational evolution.

\section{Core profiles and SIDM} \label{SIDM} 

We find that our mass profiles are adequately described by the theory of
structure formation in an \LCDM\ universe, so we can use them to bound the
strength of the dark matter particle-particle self-interaction.
\citet{sperg_stein} originally proposed SIDM as a solution to the core problem
on galaxy scales.  DSSW used numerical experiments to show that a
self-interaction cross section $\sigma_{\rm DM}$ in the range 0.5 to 5
cm$^2$ g$^{-1}$ ($\sim 1$ to 10 b(GeV/c$^2$)$^{-1}$) provided the requisite
flattening for galaxy-sized dark matter haloes.  As they have noted, a cross
section of this size is probably inconsistent with the distribution of halo
shapes in cosmological simulations (see also \citet{yswt}).  They also point
out the difficulty inherent in core profile comparisons between simulations and
observations, noting the indeterminate effect of baryon physics.

\citet{abg} showed that the mass profile for the relaxed cluster MS1358 was
inconsistent with $\sigma_{\rm DM} > 0.1$ cm$^2$ g$^{-1}$.  This was based on a
comparison with the \citet{yswt} simulations which provided an informal
relationship between cluster core size and self-interaction cross section.
\citet{markevitch} noted the need for a sample of clusters to place a robust
limit on the cross section, and derived a less restrictive limit of
1 cm$^2$ g$^{-1}$ based on a small offest between the weak lensing and X-ray
peaks of a supersonic subcluster in 1E 0657-56.

In order to gauge the dark matter self-interaction strength, we determine the
maximum core size of each cluster in the sample.  Although we see no evidence
for a constant density core in any cluster, we calculate the size of such an
unresolved core that would be consistent with the data.  We model each core as
a softened isothermal profile sphere \citep{abg}, fitting for its size and
density normalization:
\begin{equation}
\rho(r) = \frac{\sigma^2}{2\pi G(r^2+r_c^2)}
\label{eq04}
\end{equation}

\noindent where $\sigma$ is the (constant) 1D velocity dispersion and $r_c$ is
the core radius.  Since our profiles are determined in cumulative mass, rather
than density, we fit for $r_c$ and $M_0$ using
\mbox{$M_r = M_0 \, (x - \tan^{-1}x)$}, where $x=r/r_c$ and the mass scale
$M_0 = 2\sigma^2 r_c/G$.

To limit the variance of the fitted parameters which are due to a
non-hydrostatic core, we do the analysis only for the RCC subset (see
Table~\ref{t01}).  Confidence contours in core radius and mass scale are shown
in Figure~\ref{f03}.  A similar relationship obtains between core radius and
two-phase plasma to that found for core slope: two-phase emission models result
in smaller core sizes or upper limits.  As in the case of core slope, the
effect is most pronounced in A2029.  Thus the limits derived from uniphase core
models can be considered to be conservative estimates.

The core radii of the RCC subset are shown at the top in Figure~\ref{f04}.
(The corresponding core slopes for the same clusters are shown below.)  The
dashed lines show core sizes expected for three values of the dark matter
self-interaction cross section \citep{yswt}.  The DSSW cross section required
to solve the core problem -- 0.5 to 5 cm$^2$ g$^{-1}$ -- is clearly ruled out.
Using symmetrized errors, the weighted mean core size is
$\langle r_c\rangle = 35.9\pm15.4$ kpc, corresponding to self-interaction
cross section $\sigma_{\rm DM} \ll 0.1$ cm$^2$ g$^{-1}$.  We place a 99\%
upper limit of 65 kpc on the cluster core size, corresponding to just under 0.1
cm$^2$ g$^{-1}$.

A few caveats must be mentioned here.  The first is that this upper limit is
derived from a calibration with simulations which did not include the effects
of baryon physics in the core evolution.  Presumably the adiabatic contraction
of the core from radiative losses could counteract some effects of dark matter
self-interaction, and could be consistent with the observations, provided that
the contraction timescale is comparable to the mean time between dark matter
particle scatterings (to prevent cooling-induced core collapse), and that the
self-interaction is not so vigorous that gravitational core collapse proceeds
in a Hubble time.  However it is not clear that the baryons only act to hasten
core contraction.  Astrophysical processes associated with AGN or star
formation, such as jetc/ICM interactions, supernovae and stellar winds, may be
required to match observables such as the ``excess entropy'' in groups and
clusters \citet{frenk}, although the issue of feedback is beyond the scope of
this paper.

A second point is that the self-interaction cross sections in the calibration
simulations \citep{yswt} were assumed to be velocity independent.  DSSW
pointed out that a cross section which scaled as $v^{-1}$ might ameliorate the
discrepancies between SIDM cluster simulations and observations.  A functional
dependence such as $\sigma_{DM} = \sigma_0 (v/v_0)^{-\gamma}$ might reduce
the effects of self-interaction on cluster scales, although early results are
not promising (N.\ Yoshida, private communication).

Finally, if the properties of the core are in fact governed by cluster
accretion hysteresis \citep{ma}, measurements of cluster core properties will
probably tell us little until we have a sufficiently large observational sample
that we can explore the relevant physics through statistical analyses.

\section{Summary} \label{summary} 

We have presented an analysis of 10 galaxy clusters observed with the Chandra
X-ray Observatory.  We have carefully modelled their cores, utilizing a second
plasma component where required.  We have derived mass profiles for each of
them, and for the RCC subset we find that \LCDM\ provides an adequate
description of their cores.  We have placed upper limits on the size of any
constant density core in each cluster, and derive a 99\% upper limit of
just under 0.1 cm$^2$ g$^{-1} = 0.2$ b(GeV/c$^2$)$^{-1}$ for the dark matter
self-interaction cross section.





\plotone{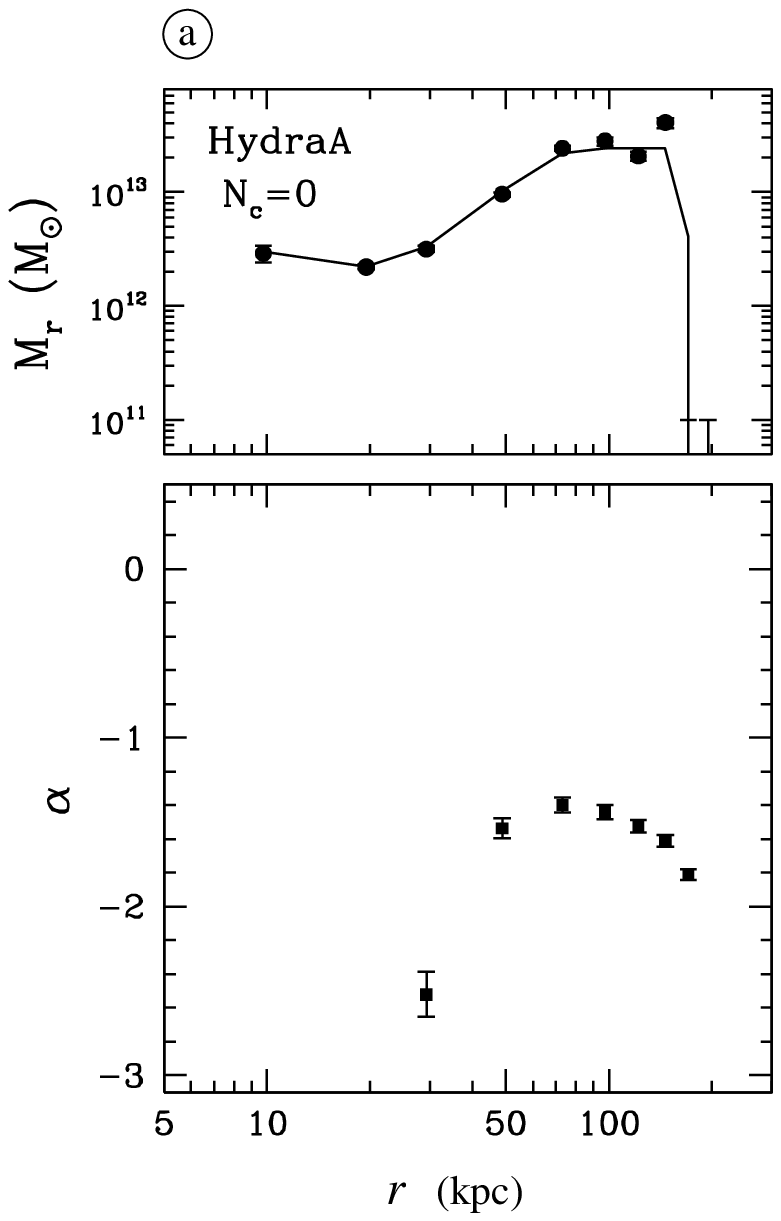}

\figcaption{Mass profiles (top) and core power law slopes (bottom) of each
cluster in the sample.  We show profiles for uniphase (left) and two-phase
(right) models of the core plasma (except for Hydra A; see text).  The
`preferred' label appears on the model with the higher MCMC significance.
\label{f01}}


\plotone{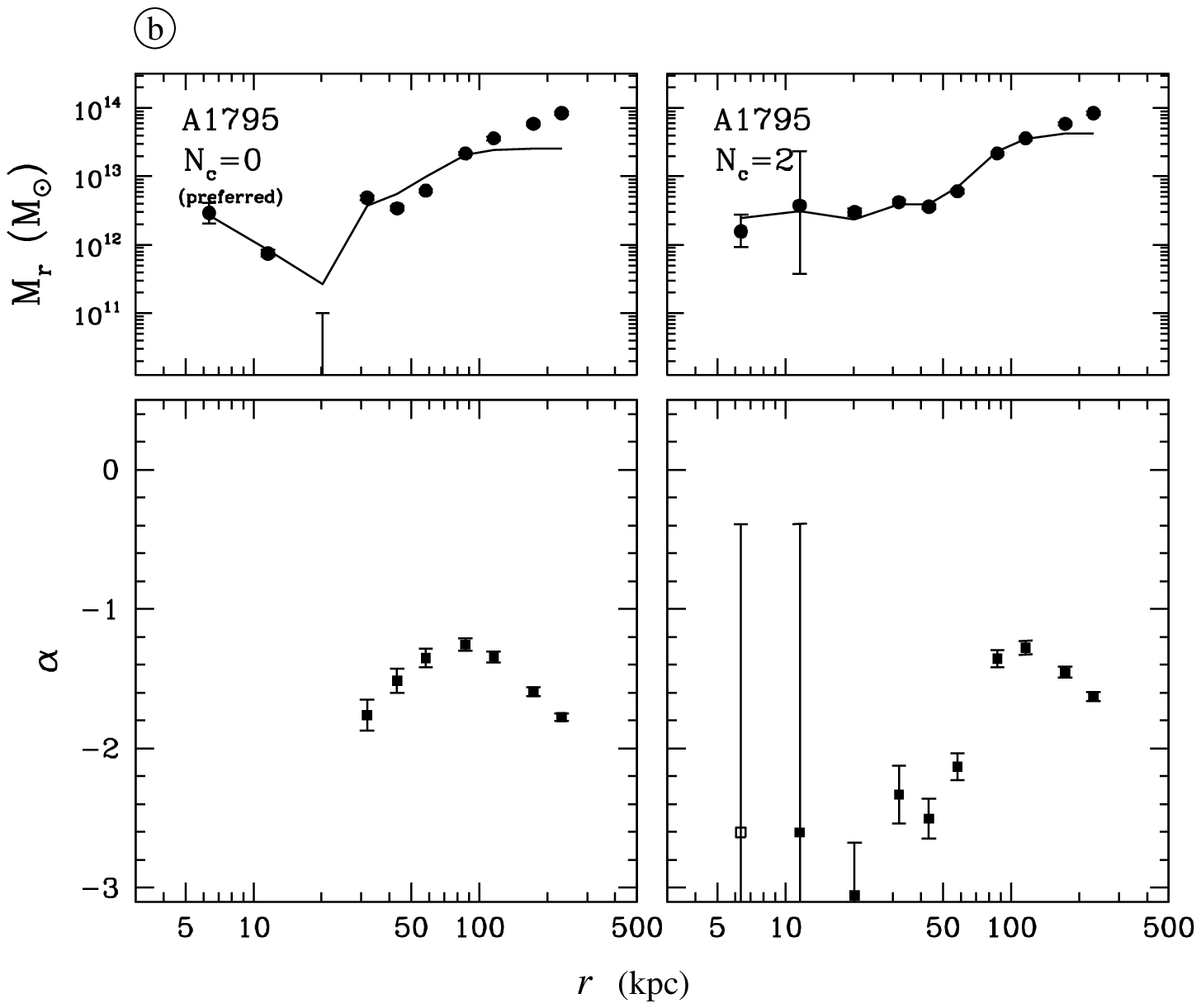}


\plotone{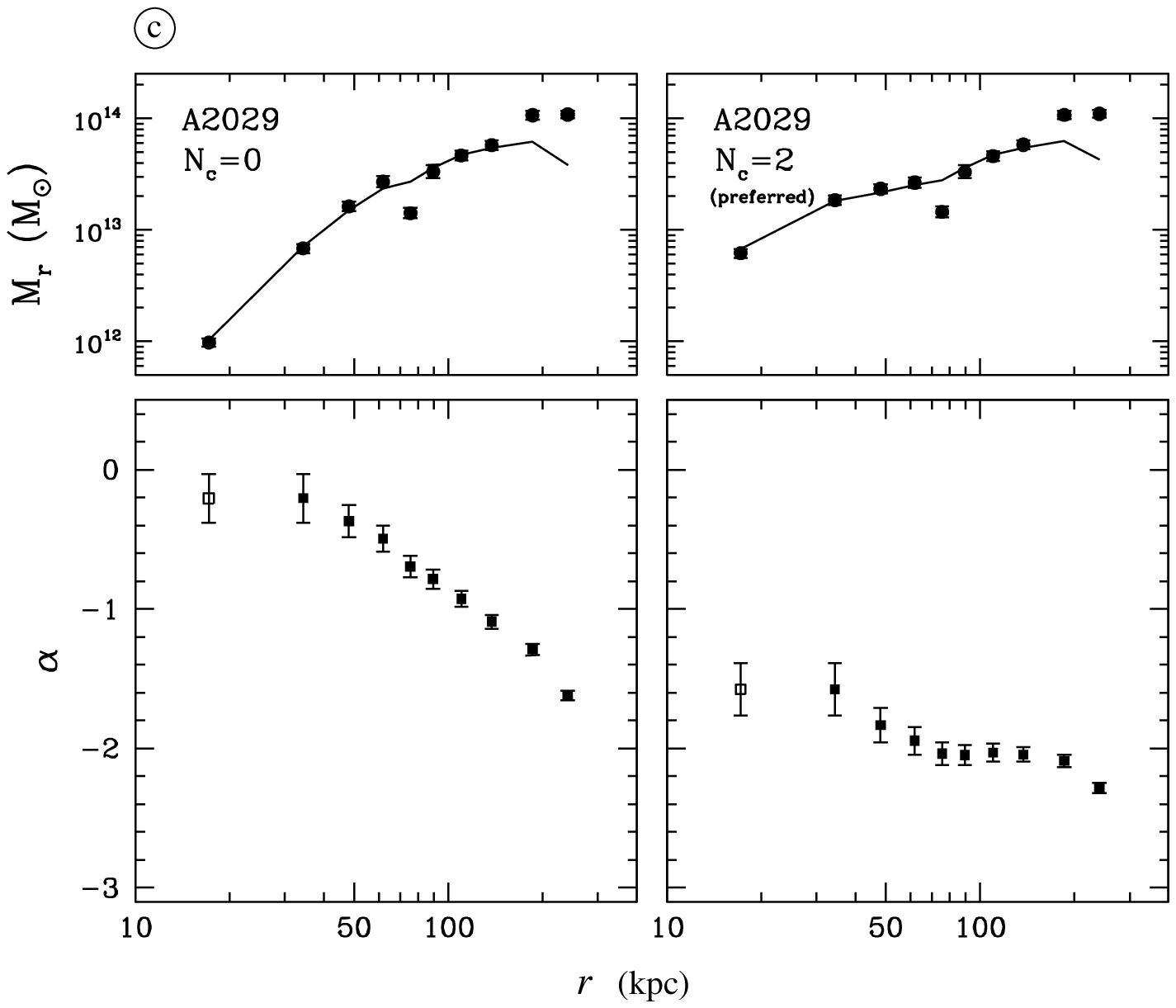}


\plotone{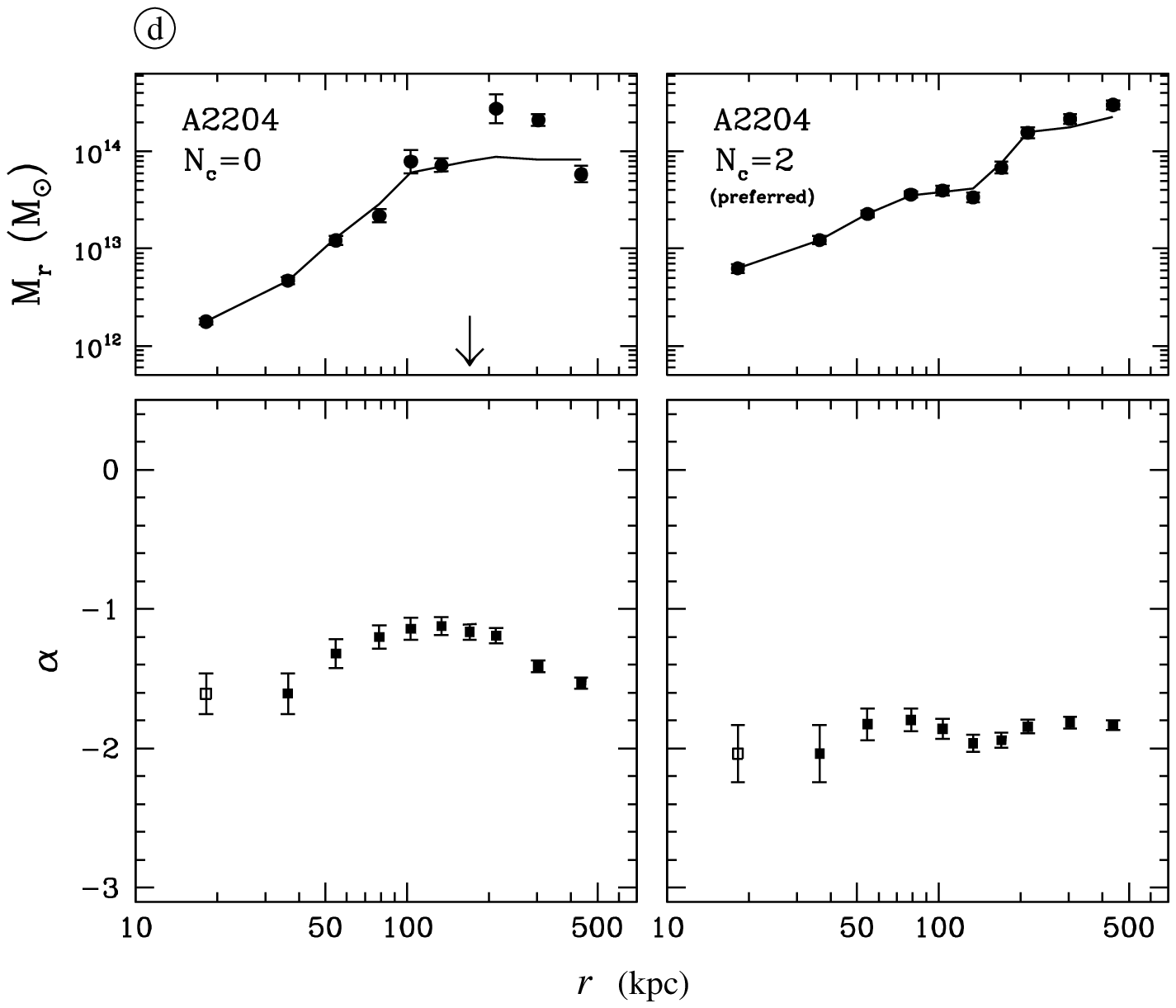}


\plotone{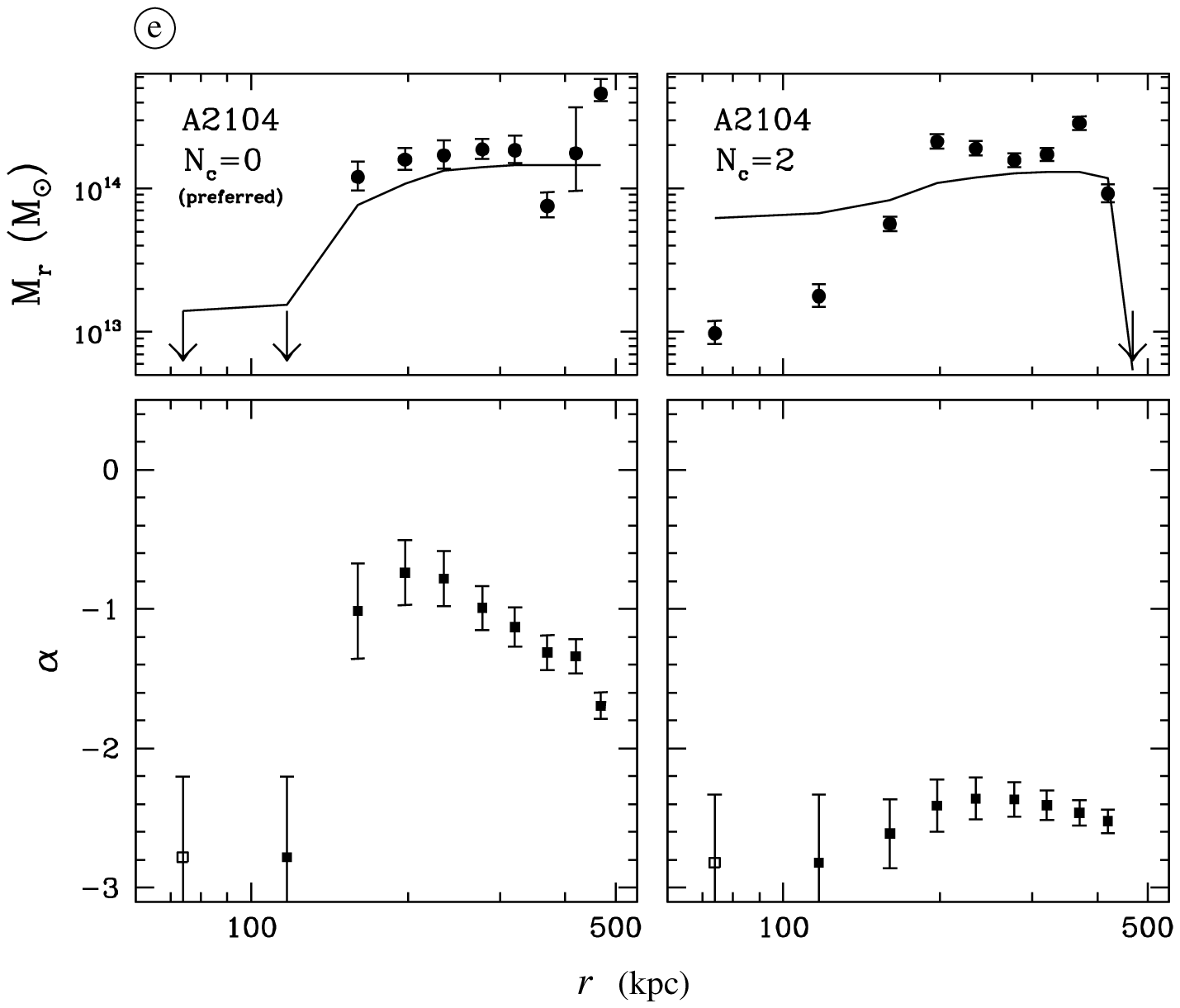}


\plotone{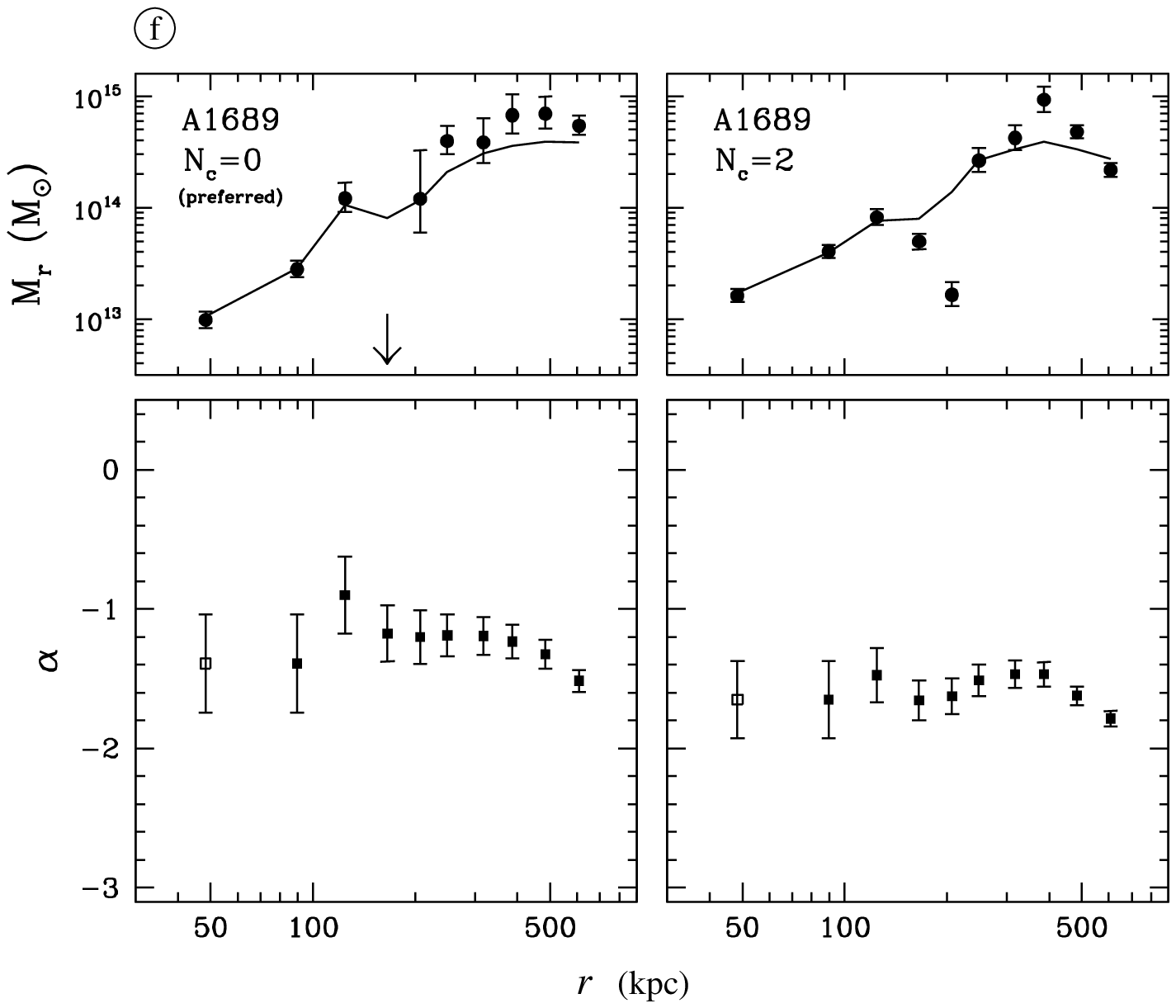}


\plotone{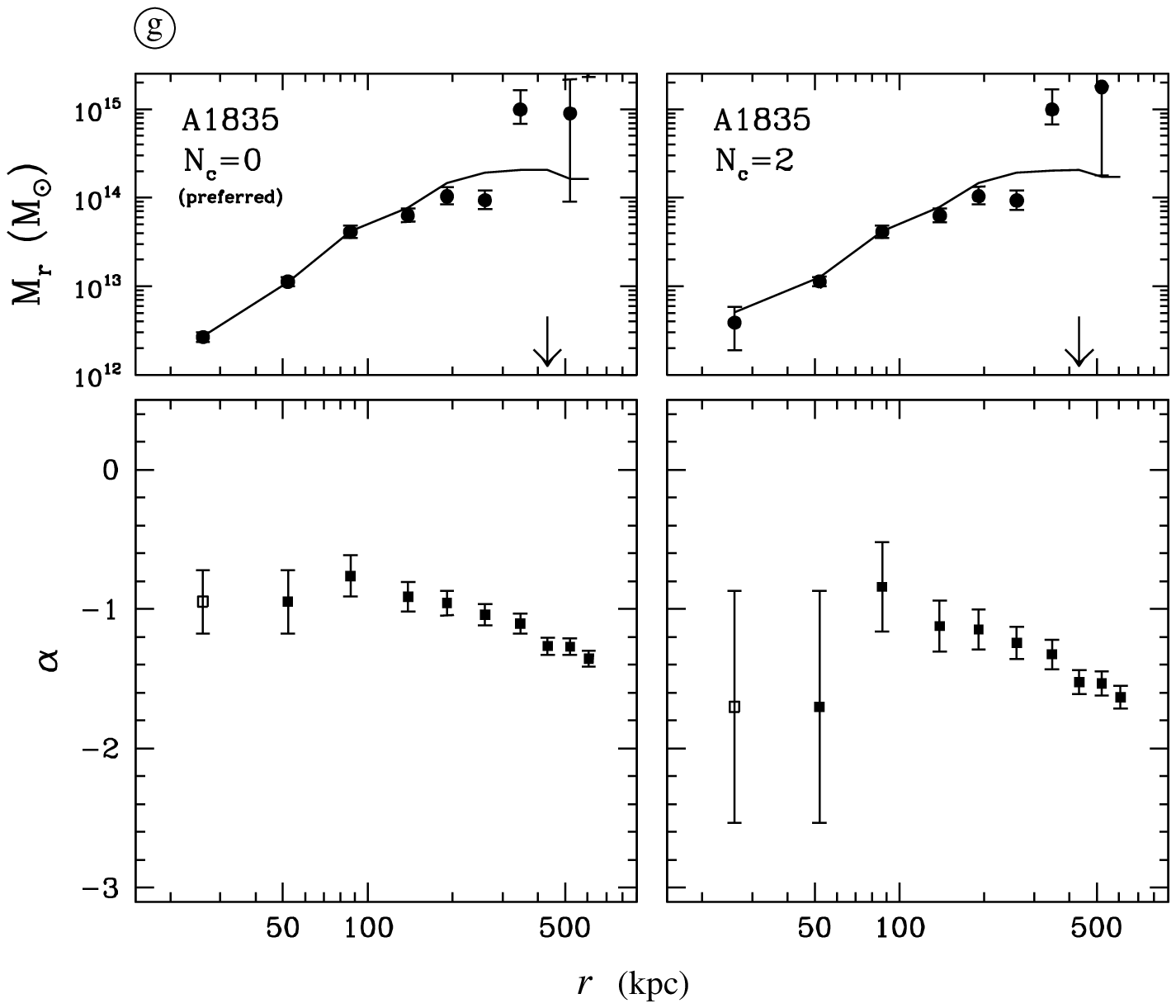}


\plotone{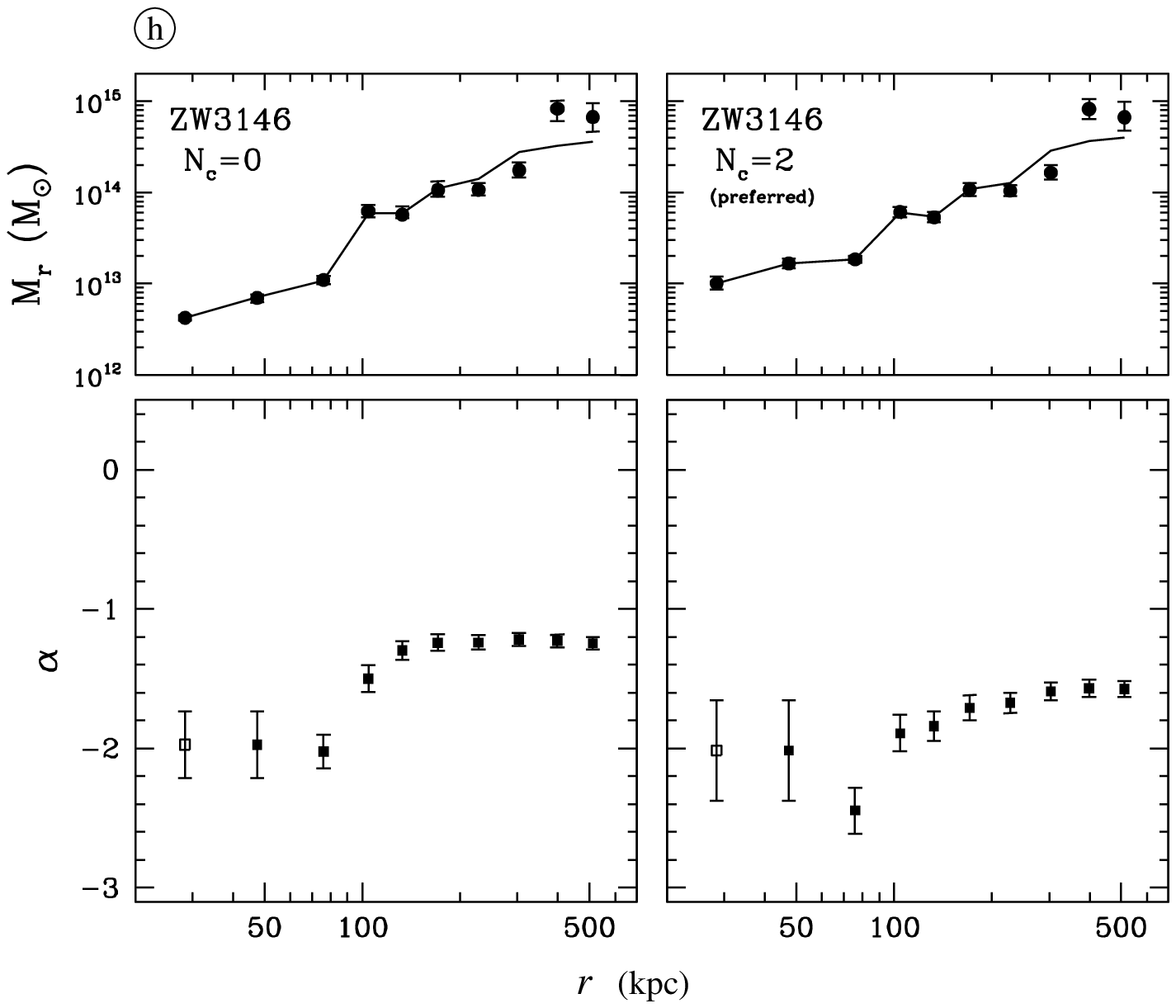}


\plotone{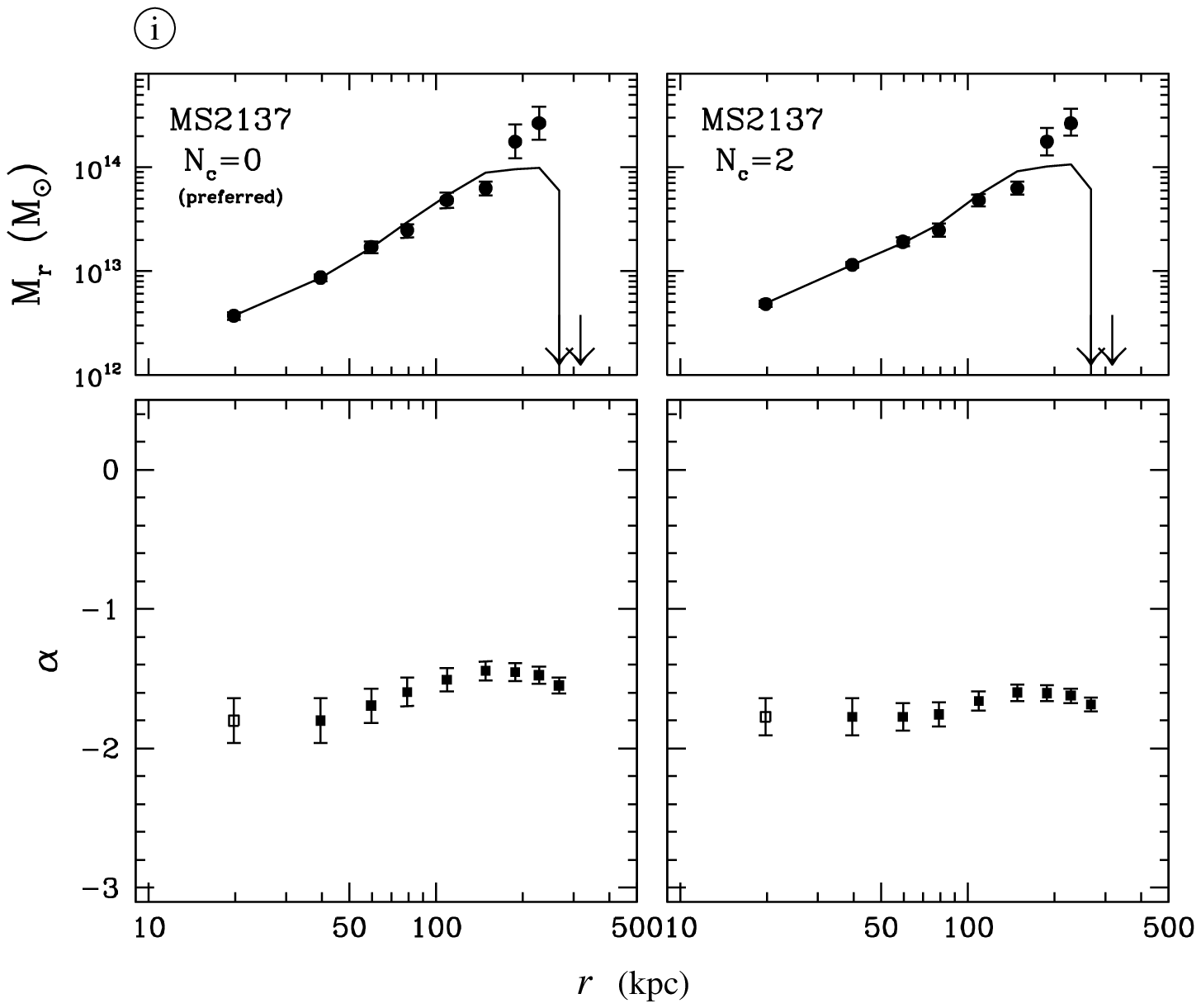}


\plotone{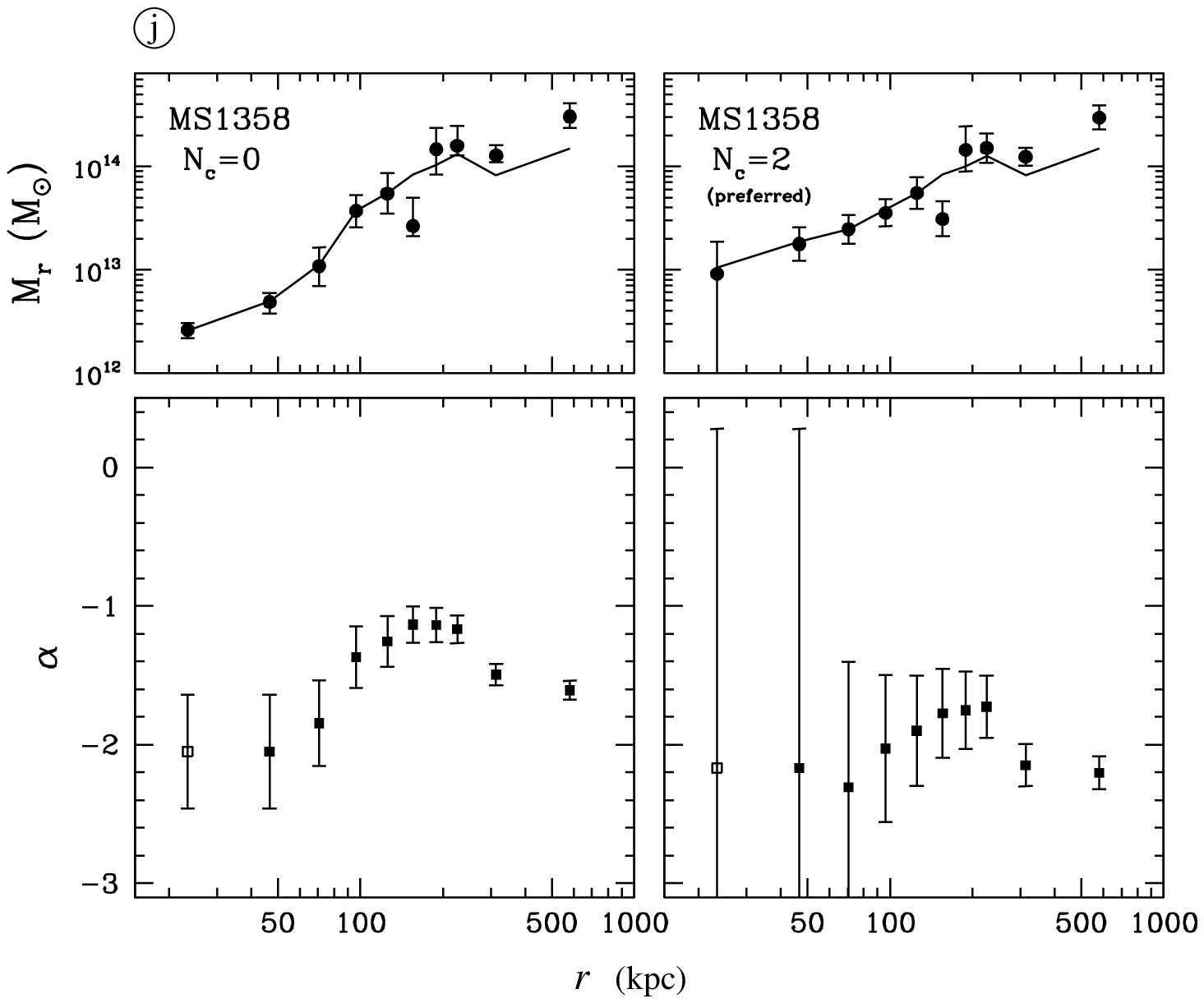}


\plotone{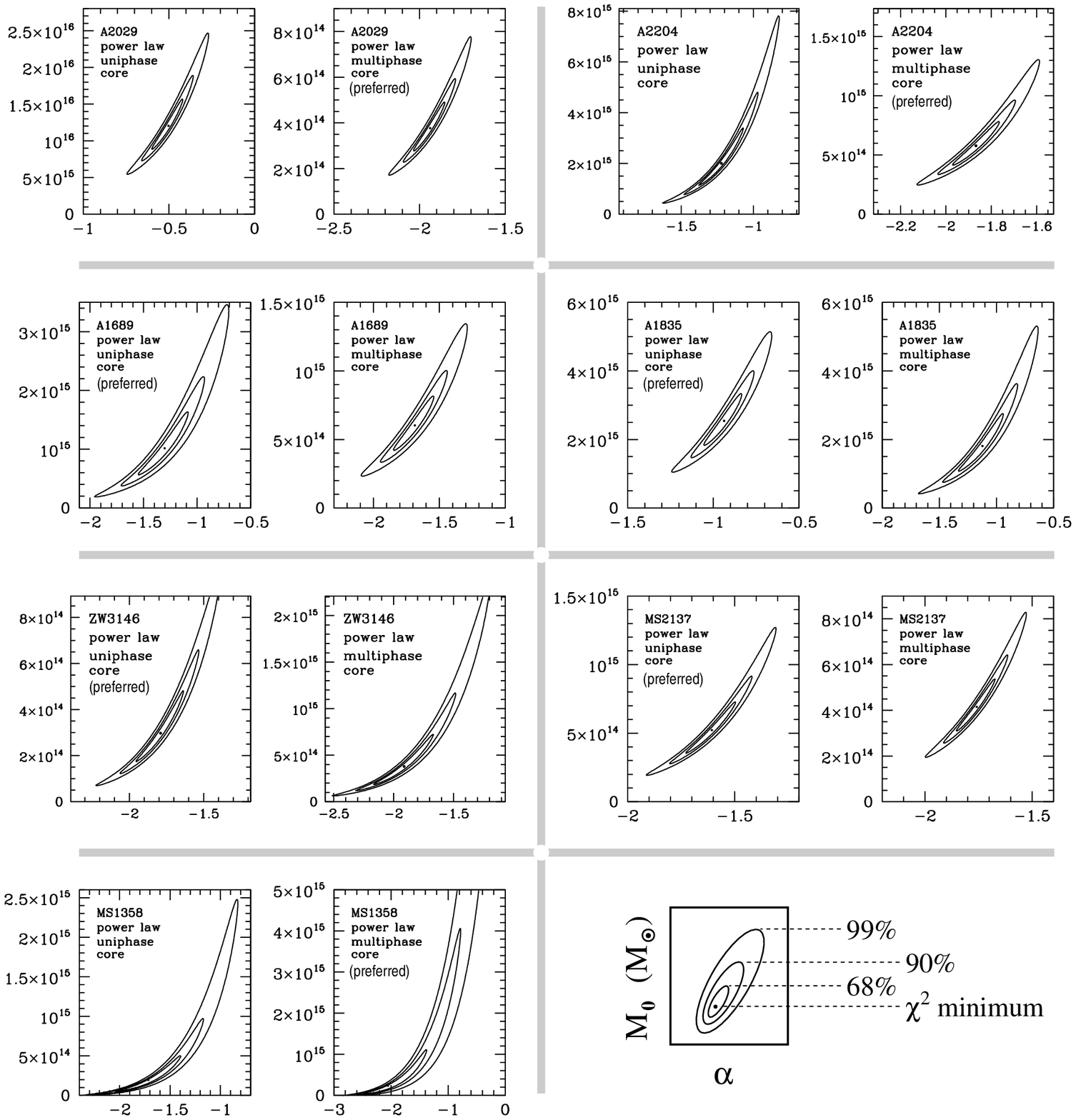}

\figcaption{Confidence contours of power law fits to the cores of the RCC
subset.  The left/right plot in each pair is for uniphase/two-phase models of
the core plasma.  The model with the higher MCMC F-test significance is labeled
`preferred'.
\label{f02}}


\plotone{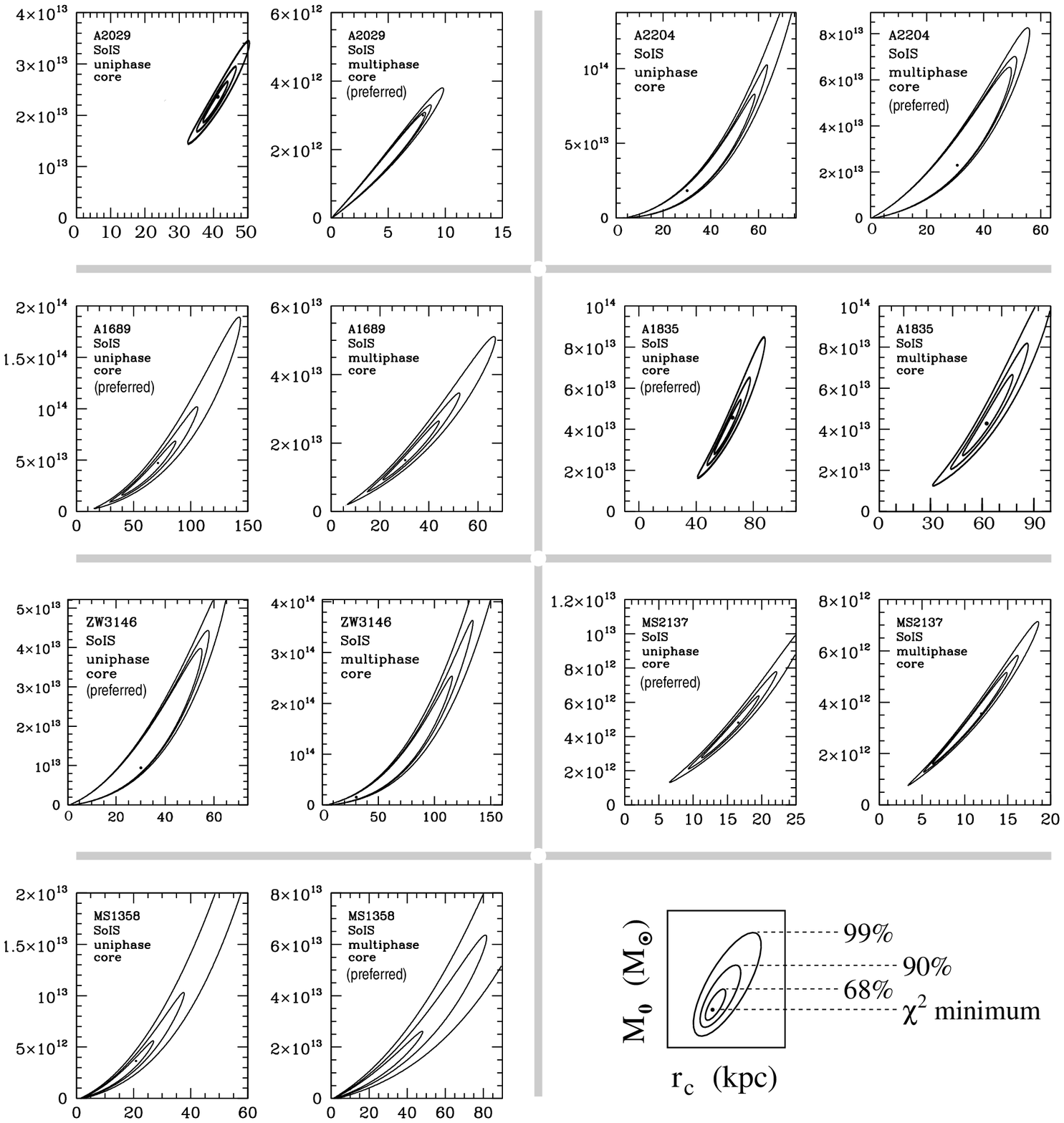}

\figcaption{Confidence contours of softened isothermal sphere fits to the cores
of the RCC subset.  The left/right plot in each pair is for uniphase/two-phase
models of the core plasma.  The model with the higher MCMC F-test significance
is labeled `preferred'.
\label{f03}}


\plotone{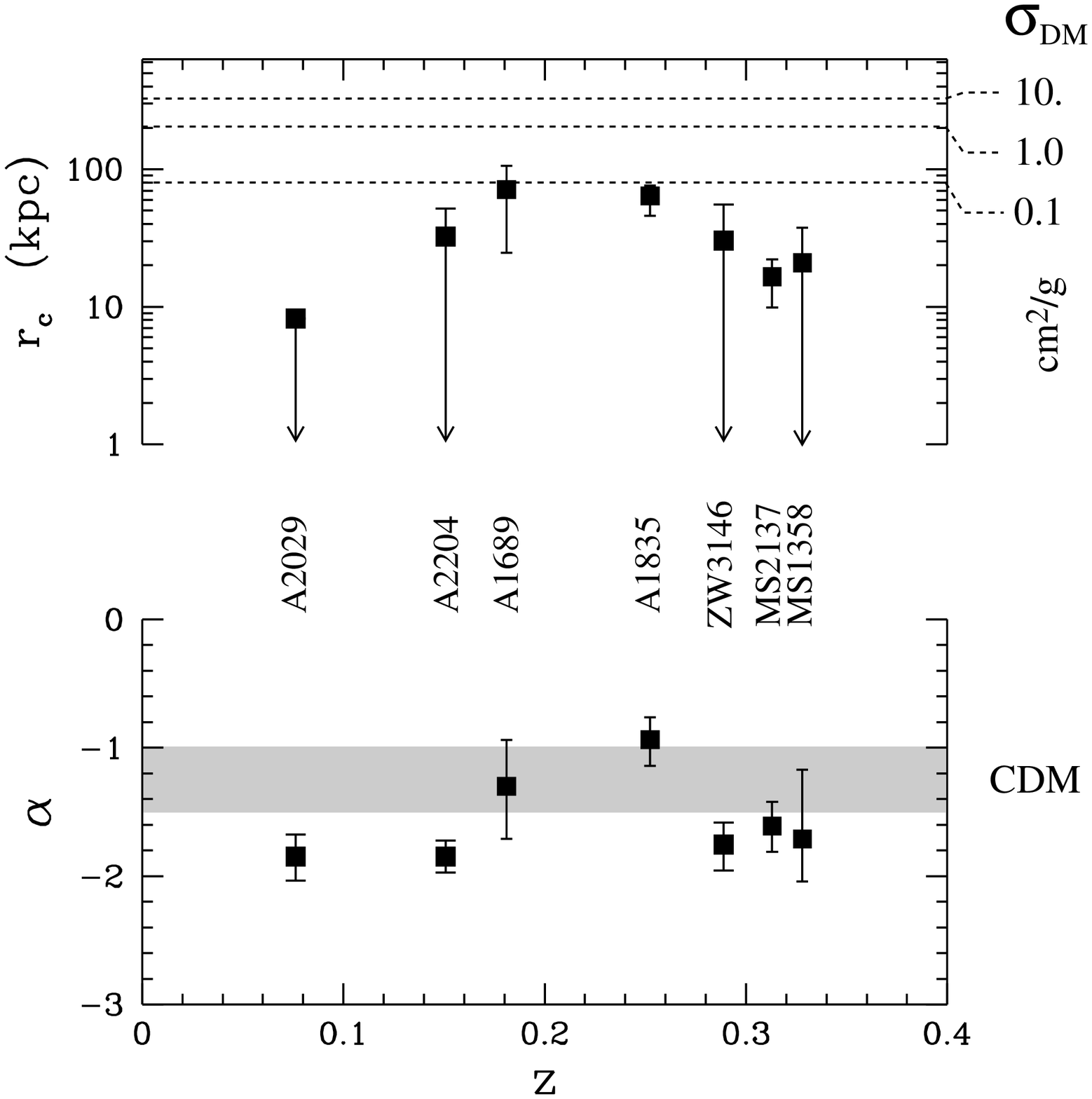}

\figcaption{Summary of the softened isothermal sphere core radii (top) and
power law slopes (bottom) for fits to the RCC subset.  The relation between the
self-interaction cross section of dark matter particles $\sigma_{DM}$ and the
cluster core radius $r_c$, taken from \citet{yswt}, is denoted by dashed lines.
The inner slope predictions of numerical CDM experiments is shown by a gray
band bracketed by the values from \citet{nfw_six,nfw_seven} at $\alpha=-1$ and
\citet{moore_b} at $\alpha=-1.5$.
\label{f04}}



\clearpage
\begin{deluxetable*}{llllc}
\tablewidth{300pt}
\tablecaption{Galaxy Cluster Sample\label{t01}}
\tablehead{
\colhead{cluster} &
\colhead{z} &
\colhead{core plasma} &
\colhead{$S$} &
\colhead{RCC?}
}
\startdata
Hydra A          &  0.0522  & uniphase   &  ---  & no   \\
Abell 1795       &  0.0631  & uniphase   & 0.823 & no   \\
Abell 2029       &  0.0765  & two-phase  & 0.998 & yes  \\
Abell 2204       &  0.1523  & two-phase  & 0.999 & yes  \\
Abell 2104       &  0.1554  & uniphase   & 0.661 & no   \\
Abell 1689       &  0.181   & uniphase   & 0.342 & yes  \\
Abell 1835       &  0.2523  & uniphase   & 0.483 & yes  \\
ZwCl 3146        &  0.2906  & two-phase  & 0.989 & yes  \\
EMSS 2137$-$2353 &  0.313   & uniphase   & 0.271 & yes  \\
EMSS 1358$+$6245 &  0.328   & two-phase  & 0.987 & yes  \\
\enddata
\end{deluxetable*}


\end{document}